\documentclass[12pt,journal,draftcls,onecolumn]{IEEEtran}
\usepackage{cite} 
\usepackage{amsmath} 
\usepackage{amsfonts,amssymb} 
\usepackage{mathrsfs} 
\usepackage{graphicx,epsfig,balance} 
\usepackage[utf8]{inputenc} 
\usepackage{CJK} 
\usepackage{multirow}
\usepackage{bm} 
\usepackage{algorithm} 
\usepackage{algorithmic}
\usepackage{extarrows} 
\usepackage{color} 
\usepackage{ragged2e} 
\usepackage{graphicx}
\usepackage{subfigure}

\begin{document}
\title{Deep Learning for Physical-Layer 5G Wireless Techniques: Opportunities, Challenges and Solutions}

\author{Hongji Huang, \emph{Member}, \emph{IEEE}, Song Guo, \emph{Senior Member}, \emph{IEEE}, \\ Guan Gui, \emph{Senior Member}, \emph{IEEE}, Zhen Yang, \emph{Member}, \emph{IEEE}, \\ Jianhua Zhang, \emph{Senior Member}, \emph{IEEE}, Hikmet Sari, \emph{Fellow}, \emph{IEEE},\\ and Fumiyuki Adachi, \emph{Life Fellow}, \emph{IEEE}
\thanks{H. Huang, G. Gui, Z. Yang, and H. Sari are with Key Lab of Broadband Wireless Communication and Sensor Network Technology (Nanjing University of Posts and Telecommunications), Ministry of Education, Nanjing 210003, China. H. Sari is also with Sequans Communications, 92700 Colombes, France (E-mails: \mbox{hongji.huang@ieee.org}, {guiguan@njupt.edu.cn}, {yangz@njupt.edu.cn}, hsari@ieee.org).}
\thanks{S. Guo is with Department of Computing, The Hong Kong Polytechnic University, Hung Hom, Kowloon, Hong Kong (E-mail: song.guo@polyu.edu.hk).}
\thanks{J. Zhang is with Beijing University of Posts and Telecommunication (BUPT), Beijing 100876, China (E-mail: jhzhang@bupt.edu.cn).}
\thanks{F. Adachi is with Wireless Signal Processing Research Group, Research Organization of Electrical Communication (ROEC), Tohoku University, Sendai 980-8577, Japan (E-mail: adachi@ecei.tohoku.ac.jp).}
}

\maketitle
\begin{abstract}
The new demands for high-reliability and ultra-high capacity wireless communication have led to extensive research into 5G communications. However, the current communication systems, which were designed on the basis of conventional communication theories, significantly restrict further performance improvements and lead to severe limitations. Recently, the emerging deep learning techniques have been recognized as a promising tool for handling the complicated communication systems, and their potential for optimizing wireless communications has been demonstrated. In this article, we first review the development of deep learning solutions for 5G communication, and then propose efficient schemes for deep learning-based 5G scenarios. Specifically, the key ideas for several important deep learning-based communication methods are presented along with the research opportunities and challenges. In particular, novel communication frameworks of non-orthogonal multiple access (NOMA), massive multiple-input multiple-output (MIMO), and millimeter wave (mmWave) are investigated, and their superior performances are demonstrated. We vision that the appealing deep learning-based wireless physical layer frameworks will bring a new direction in communication theories and that this work will move us forward along this road.
\end{abstract}

%
\IEEEpeerreviewmaketitle

\section*{Introduction}

The explosive growth of incremental data, high speed, and low latency communication scenarios has brought significant challenges to the current communication approaches, and the increasing complexity of network structures and computations degrades the performance of conventional communication systems. Recently, to improve the performance of fifth-generation (5G) communications, non-orthogonal multiple access (NOMA), massive multiple-input multiple-output (MIMO),  millimeter wave (mmWave) technologies, and other appealing techniques have been developed. However, as reported by the IBM, the explosive data will be over 40 trillion Gbits in 2020 and this is a 44-fold increase from 2009, and it is projected that the amount of connected-equipment will balloon to 50 billion. Notably, since the existing communication methods have fundamental limitations in leveraging the system structure information and handling large quantities of data, new communication theories should be established to meet the requirements of 5G systems.


Over the past several decades, much research has been devoted to developing efficient and reliable communication networks \cite{han1,han2}. Let us focus on 5G and take massive MIMO as an example. The development of channel estimation techniques has been well investigated based on eigenspace, beam space, and angle space. Unfortunately, for the eigenspace-based channel estimation methods, complicated eigen-decomposition makes them impractical for large-scale antennas systems, and their implementations are restrained in the future massive MIMO systems. Meanwhile, in the context of the beam-based channel estimation techniques such as the \emph{priori} aided (PA) channel tracking scheme \cite{pa}, single spatial support selection with the maximum amplitude for transmission leads to massive channel power leakage and disastrous performance loss. To enhance the channel estimation performance, angle space has attained great attention in recent years and angle domain-based channel estimation methods have been proposed. The existing studies assume that the angle of arrivals/departures (AoAs/AoDs) locate at discrete points in the angle domain, whereas the actual AoAs/AoDs follow continuous distribution in a practical environment. As a result, the assumption results in a power leakage problem in the angle space-based methods, which degrades the performance of the solutions. Additionally, to combine user clustering, beamforming and power allocation with successive interference cancellation (SIC), MIMO-NOMA has been proposed as a promising technique in 5G systems. However, there exist many important open research issues in MIMO-NOMA, such as trade-offs in accuracy of decoding between NOMA-weak and NOMA-strong users and the SIC effect on the achievable data rate. Although various promising techniques improve 5G system performance, complicated spatial structures and extremely difficult NP problems are not uncommon in these frameworks, and limit their practical implementation. On the other hand, in massive MIMO we have accomplished field measurements from 32 to 256 antenna elements for fully utilizing the three dimensions (3D) MIMO; how to leverage the sparsity statistics needs further investigation. Exploiting the sparsity statistics can potentially improve massive MIMO system performance, and it is worth to investigate whether joint optimization of various fundamental elements in the communication system is possible and can result in system performance improvement.

Inspired by the recent advances in deep learning paradigm, deep learning-based wireless communication techniques have aroused considerable interest among the academic and industrial communities. Different from previous works, deep learning-based communication scenarios may have a high potential in the areas such as multiple access, channel estimation, and performance optimization for the following reasons.

\begin{enumerate}
  \item Most existing communication systems are designed block by block, which means that these systems comprise multiple modules such as transmitter and encoder. For this block-based architecture, many techniques have been developed for optimizing the performance of each block, but not that of the whole system. Recently, several new studies \cite{csi,li,amp,noma-g,mimo,iot,jin2,mmw} have demonstrated that end-to-end optimization, i. e., optimizing a whole communication system, is superior to the block-based optimization. Deep learning provides a state-of-the-art tool for achieving end-to-end performance maximization, with the potential of deep neural networks (DNNs) to accurately model communication systems.
  \item In 5G systems, complex and large-scale scenarios lead to various communication links with rapidly changing channel conditions. A large number of models, such as the joint spatial division multiplexing (JSDM)-based massive MIMO model relies heavily on channel state information (CSI), and their performance can deteriorate in imperfect and nonlinear scenarios \cite{csi}. Acquiring accurate CSI of time-varying channels is essential to the system performance. By taking advantages of deep learning techniques, it is possible that the communication system can learn emergent channel models and be trained to adapt to new channel conditions.
  \item Massive concurrent architectures with distributed memory, such as in popular graphical processing units (GPUs), have shown to be energy efficient and perform well, resulting in impressive computational throughput when leveraged by concurrent algorithms \cite{csi}. Deep learning-based schemes are suitable for running on GPUs to fully utilize such parallel hardware because of the DNN structure.
  \item The emerging 5G systems require fast and effective signal processing algorithms to deal with huge data and rich content scenarios. However, low-resolution analog-to-digital converters used in current frameworks cause unexpected nonlinear imperfections, which calls for rigorously robust processing methods with high computational complexity \cite{ch}. Fortunately, deep learning based methods can help handle massive data and rapidly changing scenes based on parallel processing architectures.
\end{enumerate}

Motivated by the preceding considerations, we conduct an investigation of deep learning for wireless physical layer. To support new demands in future communication scenarios, our research emphasizes 5G systems, and the following three potentially disruptive methods may bring about technical advancements, to make the 5G networks easily for practical implementations.

\begin{enumerate}
  \item \textbf{Deep learning-based NOMA:} NOMA is a 5G technique due to its potentials for boosting spectral efficiency and system capacity; however, user mobility leads to highly complex channel conditions in each link. The performance of NOMA-based frameworks depends on the CSI, because interference cancellation requires accurate knowledge of the very complex channel conditions. We incorporate deep learning into the NOMA framework and corroborate its superior performance.
  \item \textbf{Deep learning-based massive MIMO:} As perfect CSI has great effect on potential gains of massive MIMO, it would make sense to investigate high-accuracy channel estimation issue for acquiring perfect CSI. We consider frameworks that adopt deep learning techniques into the massive MIMO systems for direction of arrivals (DOA)  and channel estimations by fully utilizing the spatial information. Furthermore, it is worth pointing out that the DNN is a good candidate to conduct accurate CSI reconstruction for super-performance channel estimation of the massive MIMO.
  \item \textbf{Deep learning-based mmWave:} MmWave communication has a ten-fold increase in bandwidth compared with the current wireless networks, but mmWave-based scenarios lead to ultra-high power consumption and limited link gains in correlated channels \cite{3mm}. Here, we employ deep learning in an mmWave massive MIMO system and show its excellent performance in terms of hybrid precoding.
\end{enumerate}

\section*{Overview of Deep Learning for Wireless Communication}

In general, DNN has become one of the most universal generative models. As stated by the popular \emph{universal approximation theorem}, by employing the multilayer perceptron (MLP) mean, a feed-forward neural network with a single hidden layer is capable of approximating continuous functions on compact subsets of $R^n$, where $R^n$ represents the field of real numbers with $n$ dimension. Given sufficient samples, the DNN is able to extract important features from network inputs and realize end-to-end learning for predicting or regressing. For instance, if the DNN is trained based on a great number of images comprising the cats, it can capture essential features such as nose and fix these significant features into the next layer of the network. After training the network, it enables to recognize the cats if we provide some images including the cats. In recent years, much research has been devoted to studying deep learning-based wireless communication, and its efficiency has been demonstrated in several results, including in the following studies.

\begin{figure*}[!t]
\centering
\includegraphics[width=155mm]{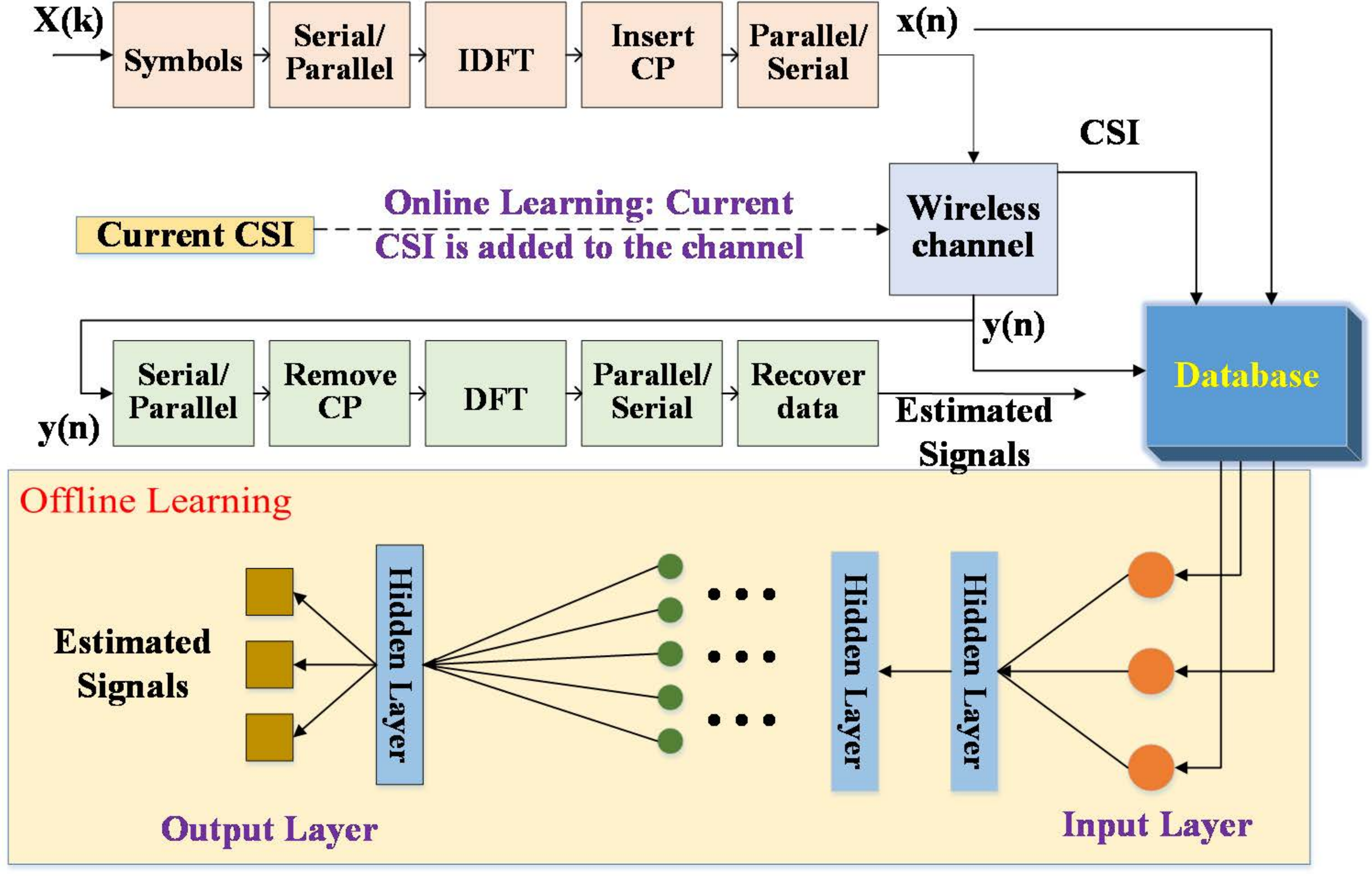}\\
\caption{DNN architecture for channel estimation in OFDM systems. Here, $x(n)$ and $y(n)$ denote the transmitted signal and the received signal, respectively. Also, $X(k)$ is defined as the DFT of $x(n)$.}
\label{ce}
\end{figure*}

\begin{itemize}
  \item \textbf{Deep learning for channel estimation:} For the sake of achieving high-resolution channel estimation, a DNN for channel estimation is developed for orthogonal frequency-division multiplexing (OFDM) systems \cite{li}, as presented in Fig. \ref{ce}. First, pilot symbols are inserted into the transmitted symbols which are transformed into parallel data flows; then the inverse discrete Fourier transform (IDFT) converts the signals $X(k)$ from the frequency domain into the time domain. Thereafter, to remove the inter-symbol interference (ISI), a cyclic prefix (CP) is inserted into the data stream. Subsequently, after training the model under different channel conditions, the output generated by the DNN recovers the input symbols $x(n)$ without requiring explicit channel detection. To be specific, in this procedure, we feed the transmitted symbols and received OFDM signals into the DNN and train the DNN to minimize the difference between the input and the output of the network. As shown by simulations, this method is more efficient than those in previous studies, since it achieves better performance by adding fewer pilot symbols and removing the CP.
  \item \textbf{Deep learning for encoding and decoding:} In the DNN decoder \cite{csi}, a novel neural network (NN) with multiple dense layers optimized by batch normalization is constructed, and the transmitted signals are encoded as a one-hot vector. The signals transmitted over the wireless channel are modeled as a noise layer, and are conveyed to the NN-based receiver.
      Next, this model is trained by a stochastic gradient descent (SGD) algorithm, and the output signals with the highest probability are the decoded messages, i.e., $k$ bits comprise $2^k$ messages. Here, the cross-entropy loss function is used as to process the model. The simulation results show that the deep learning-based encoding and decoding method can infinitely approach the performance of the Hamming code, in the absence of requiring encoder and decoder functions.
  \item \textbf{Deep learning-based signal classification:} Automatic modulation classification (AMC) is an appealing technique for environment identification and transmitter identification. Inspired by the strengths of the long short term memory (LSTM), deep learning-based signal classification is developed \cite{amc}. The framework is trained to classify 11 typical modulation types in the LSTM based on input signals in polar coordinates form. The signal classification performance is demonstrated at a high signal-to-noise ratio (SNR). By contrast, in the context of low SNR, Convolutional Neural Network (CNN) is a potential candidate to achieve high-accuracy signal classification by using the powerful learned filters. This framework can be divided into pre-training stage and fine-tuning stage. Hence, a mixture model-based framework (e.g., a system comprises the LSTM and the CNN modules) is regarded as a solution to realize high performance in terms of AMC with various SNR regions.
  \item \textbf{Deep learning-based MIMO detection:} Detection Network (DetNet), as an emerging deep learning-based framework by unfolding a predicted gradient descent mean, can be applied to multiple models by conducting single training. For example, MIMO detection can be achieved by using typical searching algorithms, but its performance degrades due to the worst case computational complexity. For both time-invariant channel and time-varying channel scenarios, we construct a detection architecture and derive an optimal estimation $\widehat{\mathbf{x}}_{\theta}(\mathbf H, \mathbf y)$. Here, $\mathbf x$ represents an unknown vector of binary symbols, while $\mathbf H$ and $\mathbf y$ represent the channel matrix vector and the received vector, respectively. The task is to obtain the parameter $\theta$ by minimizing the loss function according to the least square principle. It is noted that the performance of the MIMO detection is related to the choice of loss function and the developed NN. Simulation results verify that it achieves superior performance in terms of bit error ratio (BER) as compared with several existing works, including the zero forcing, approximate message passing and semidefinite relaxation approaches.
\end{itemize}

\section*{Deep Learning for 5G: An Alternative Approach}

\subsection*{\centering{Deep Learning-based NOMA Framework}}
\label{sec2}

\begin{figure}
  \centering
  \includegraphics[width=87mm]{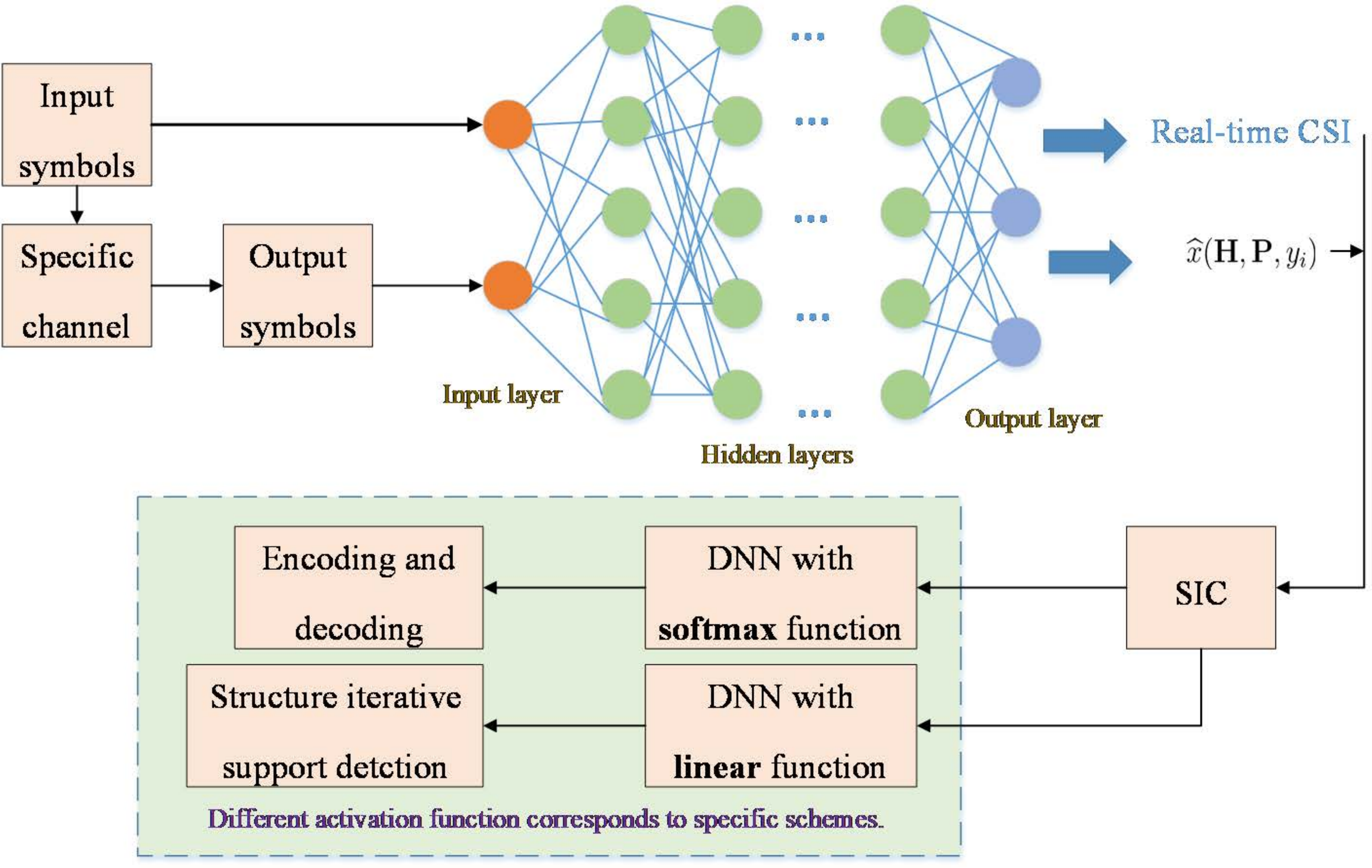}\\
  \caption{Deep learning-based NOMA system for data detection, encoding, and decoding.}
  \label{fig_sys}
\end{figure}

Consider a typical NOMA system in which a base station (BS) serves multiple users and various channel conditions exist in each communication link. For the sake of estimating real-time CSI at each user and addressing optimization issues in the fields of signal detection and encoding, we design a DNN framework, as illustrated in Fig. \ref{fig_sys}. Here, different from the conventional block-based system, we regard the whole developed NOMA system as a blackbox and the DNN is employed for approximating the whole NOMA system, which comprises the BS, wireless channels, and all the users (signals receivers), etc.. The DNN with 18 layers is used to process an enormous quantity of data including original transmitted signals and channel vectors \cite{noma-g}. There are 16 hidden layers with multiple neurons that are used to perform training and recognition. The Rectified Linear Unit (ReLU) function is introduced to activate all the hidden layers. The input layer is a fully connected layer with 128 neurons. The original transmitted signal vectors are fed into the input layer and then their essential features are broadcasted to the first hidden layer. The second and third hidden layers for encoding are dense layers that comprise 400 neurons and 256 neurons, respectively. Then, we design one of the hidden layers as a noise layer including 256 neurons to represent the wireless channel which is polluted by the AWGN. Next, there are two hidden layers that are regarded as a decoder for transferring the vectors into the codewords. In addition, a linear function is adopted in the output layer, and we obtain the output signals based on the NOMA system.

The deep learning-based NOMA framework requires a learning policy for training the DNN framework. Current techniques allow us to design a simulator for simulating different channel conditions such as fast fading and slow fading channels scenarios since typical channel models that can depict the realistic channels well have been derived. In each simulation, we generate transmitted data sequences, and various channel models are then used to generate the CSI according to the NOMA principles. The transmitted data and the output data obtained at the output layer of the DNN are collected as a set, namely, the training data set. To deploy the power allocation policy automatically and efficiently, examples with different power levels for multiple users are allotted according to the power allocation factor $\alpha$ in the training phase. Here, the input of the DNN is the power allocation vectors based on traditional methods, the transmitted signals, and the $\alpha$, whereas the output of the network is the best power allocation policies. After training the DNN with the SGD based on the loss function, we obtain the best power allocation policy after hundreds. Afterwards, the online learning approach is employed to process the well-trained DNN framework based on current input signals and real-time channel characteristics, contributing to achieve real-time channel detection. It is worth stating that the power allocation issue can be realized automatically for any amount of users or quality of the CSI, while conventional NOMA-based power allocation policies fail to extract the channel conditions adaptively and they can only be deployed in a specific scenario.

To address the multiple-user detection issue and avoid the users transmitting signals randomly that leads to signaling overhead and transmission latency of uplink NOMA network, where user activity is required to be detected at the BS, we investigate user activity and data detection issues in which users are in a time slot including several continuous time slots (one main advantage in 5G). Using the developed DNN framework with the aids of the proposed learning policy, the computational complexity and data detection performance have been enhanced. Initially, we denote $\mathbf{y}^{[j]}$ and $\mathbf{x}^{[j]}$ as received signal and structural sparsity in the $j$-th time slot, respectively. Additionally, $\mathbf{H}_s^{[j]} \in \mathbb{C}^{n\times S}$ represents the channel matrix, where $S$ and $n$ denote the amount of OFDM subcarriers and users, respectively.

\begin{figure}
  \centering
  \includegraphics[width=87mm]{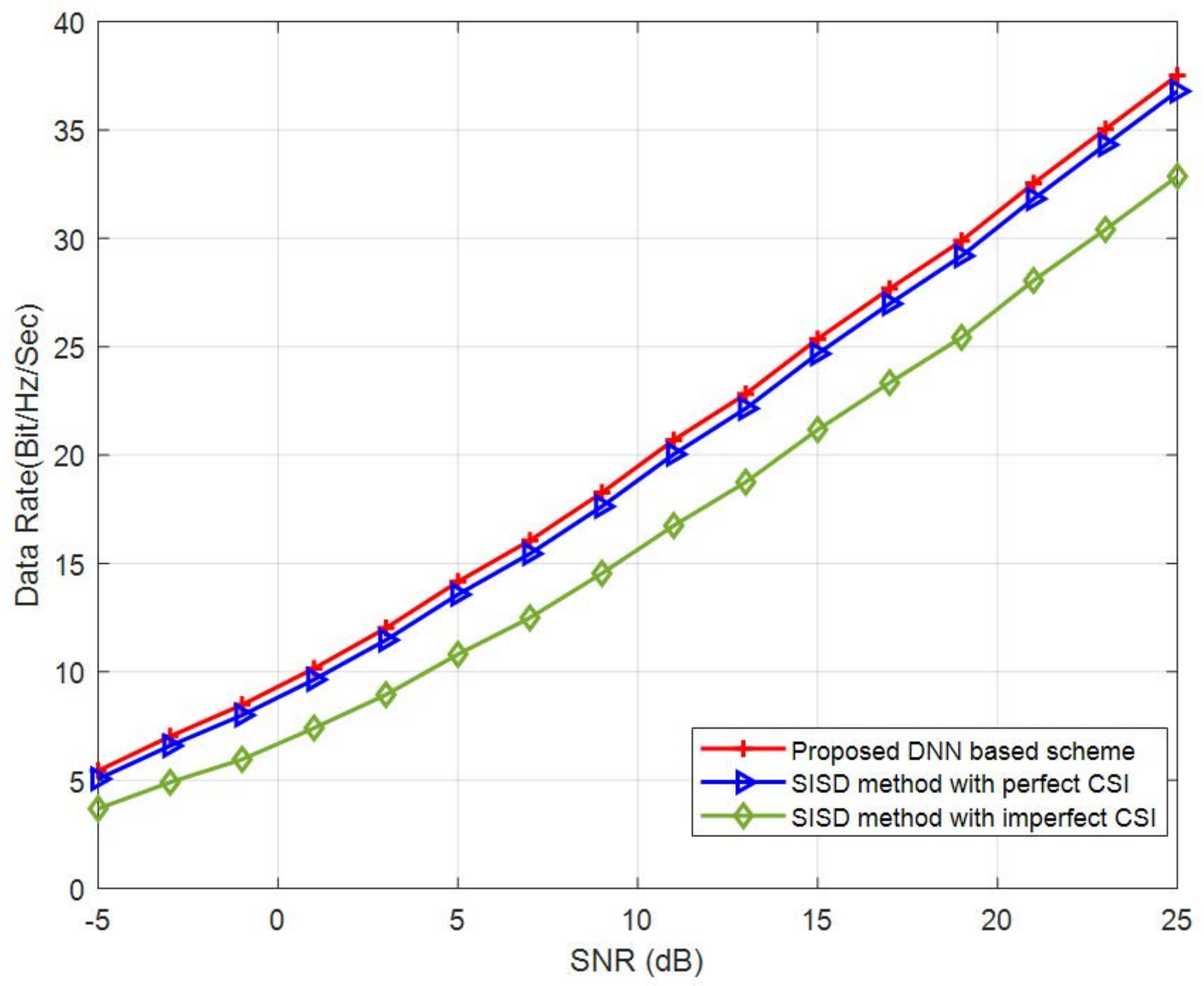}\\
  \caption{Data rate for the deep learning-based mean and the SISD method at different CSI.}
  \label{fig_noma}
\end{figure}

Then, to apply the user activity and data detection technique in the DNN, we first perform an environment simulator to emulate the wireless channel and corrupt artificially created noise into the channel. Considering this relation to be a function, the DNN framework is adopted to minimize the difference between $\mathbf{H}_s^{[j]} \mathbf{x}^{[j]}$ and $\mathbf{y}^{[j]}$. It is noted that each iteration of user activity and data detection is equal to a constant mapping transformation in the DNN. Due to the factor that the data-driven approach is an emerging technique for wireless communication society, the common dataset for model training is in general not available. As a result, simulation-based data generation has been widely used recently, e.g., \cite{li,amp,noma-g,mimo,jin2,mmw}, and witnessed promising results. This paper adopts the same approach. In our developed framework, we generate transmitted data sequences randomly and obtain corresponding NOMA frames with pilot symbols. Then, the channel vectors are obtained after the NOMA frames being fed into the given channel model. Here, the DNN is trained by 250,000 batches of training dataset in which each batch has 24 data sequences. Also, the weights and bias of the DNN are initialized by the well-known Xavier mean. As depicted in Fig. \ref{fig_noma}, the proposed DNN-based approach outperforms the support detection (SISD) algorithm regardless of the CSI. This improvement is supported by the fact that the DNN-based method can obtain a real-time CSI of the link between the BS and each user with high precision, which implies that deep learning is a promising approach for resolving the power allocation and data detection issues in NOMA.


Encoding and decoding are two key parts of a communication system. In the DNN-based framework, we consider the whole system to be an autoencoder, in which the encoding and decoding procedures can be jointly processed through neurons interacting. To simplify the description, we introduce an indicator function which is formulated as $\mathbf 1_{x_{i,1}}$. Here, $x_{i,1}$ denotes the transmitted signal of the $i$-th user. With regard to this function, all symbols are zero except $x_{i,1}$, while $x_{i,1} = 1$. Additionally, a softmax function is implemented at the output layer of the DNN to work out a probability distribution over all the possible codewords; the estimated symbol $\widehat{x}_{i,1}$ with the highest probability is the decoded signal. To evaluate the performance of the deep learning-based NOMA method, the block error ratio (BLER) is introduced and computer simulation has verified that the proposed scheme achieves a lower BLER compared to the hard decision method and the original data. Thus, combined with the results of the data detection analysis, we argue that incorporating deep learning into NOMA is an effective and reliable strategy.

\subsection*{\centering{Deep Learning-based Massive MIMO}}

Although massive MIMO is a possible strategy for future communication networks to maximize spectrum efficiency, a very large spatial complexity of the system reduces its performance. Concentrating on the direction of arrivals (DOA) estimation and channel estimation issues, we apply deep learning to a massive MIMO system, and the feasibility of this novel framework is verified using extensive simulations \cite{mimo}.

\begin{figure*}[htbp]
\centering
\subfigure[MSE via SNR performance of the proposed deep learning based-approach, SBEM method, PA channel tracking mean \cite{pa}, ADMA user scheduling method, and MUSIC approach.]{
\includegraphics[width=87mm]{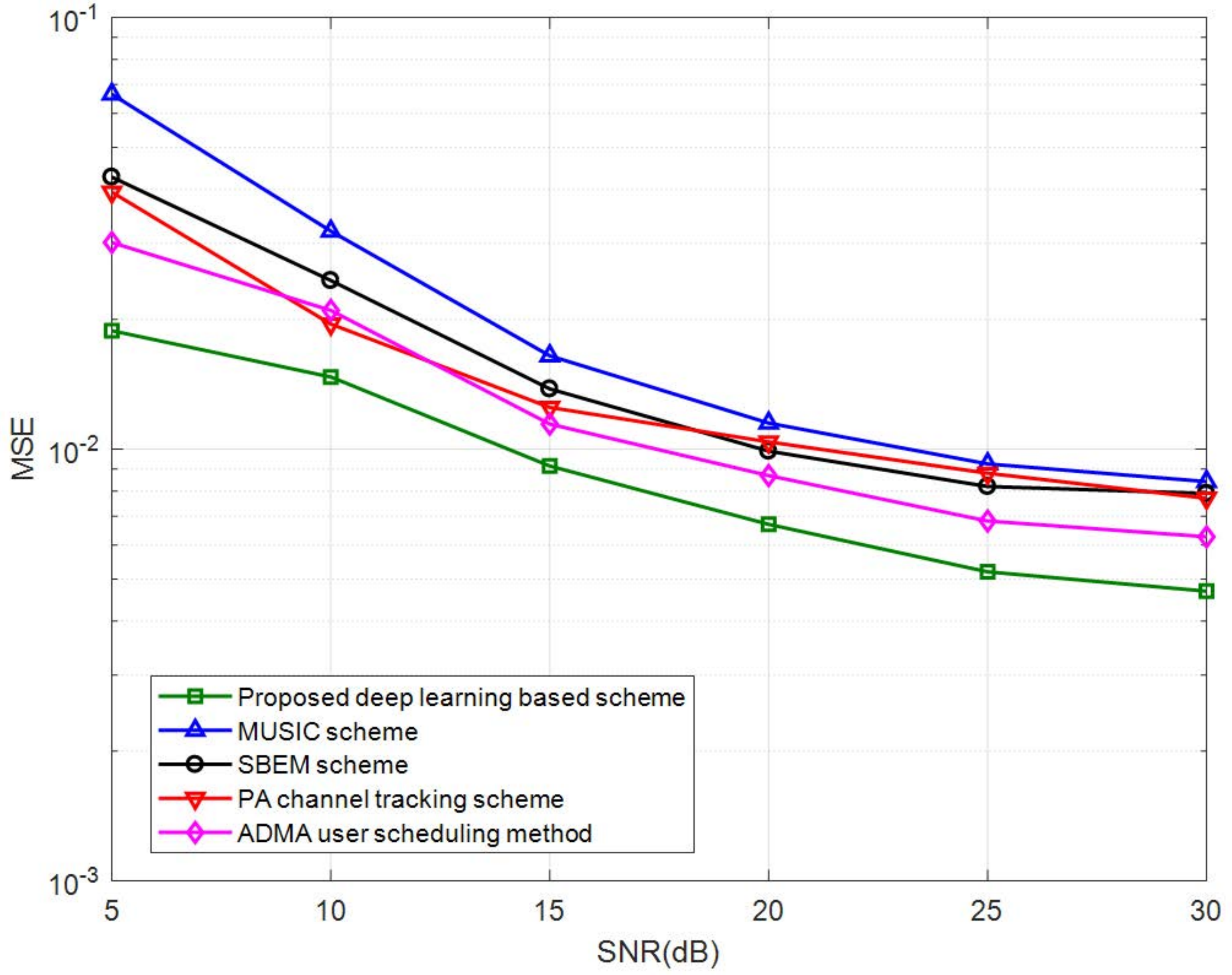}
\label{com}
}
\subfigure[BER via SNR performance comparison of the proposed deep learning-based approach, the SBEM method, and the JSDM method.]{
\includegraphics[width=87mm]{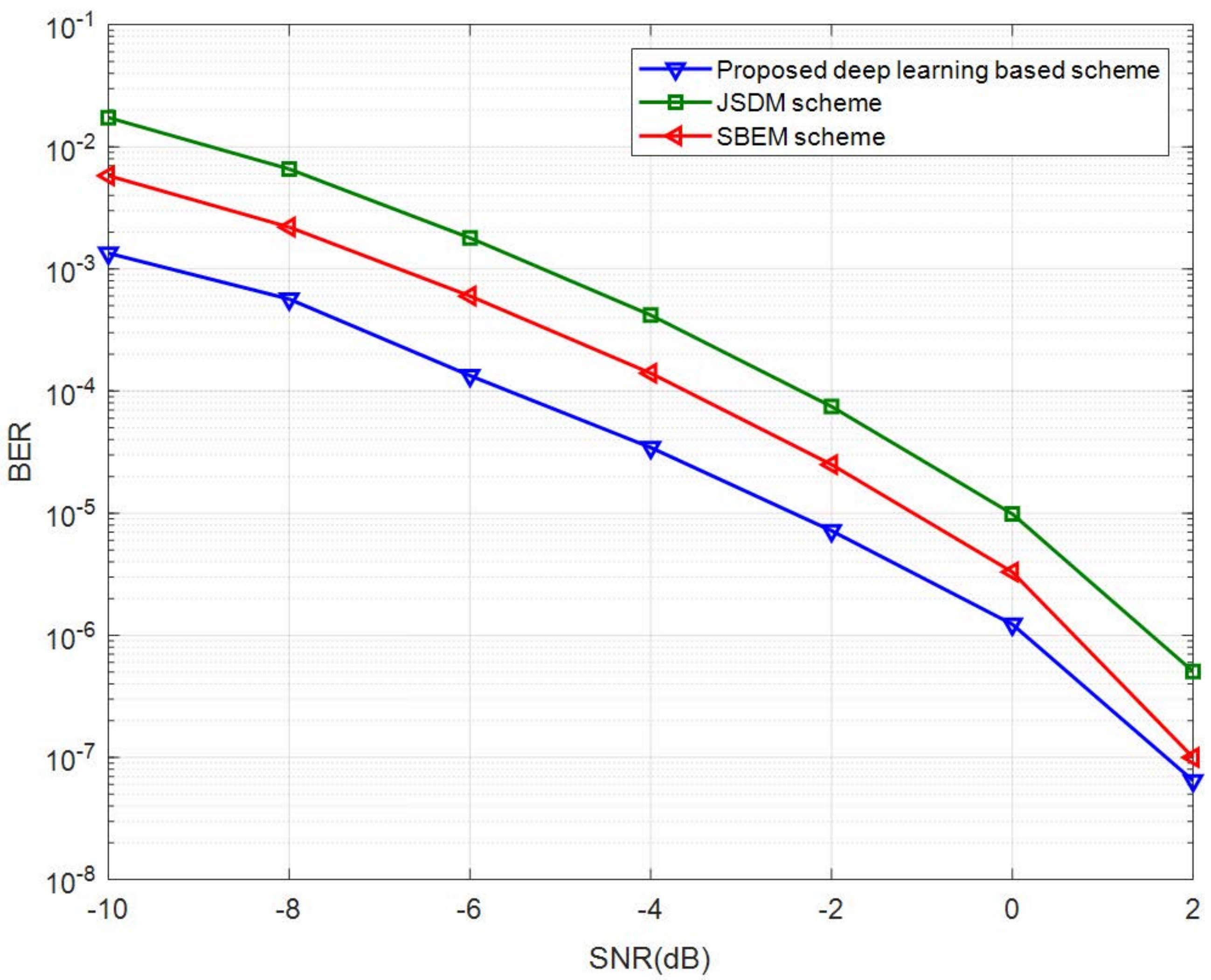}
\label{channel}
}
\centering
\caption{Deep learning-based massive MIMO system.}
\end{figure*}



With respect to the proposed deep learning-based framework, we regard the uplink massive MIMO system that is composed of a BS with multiple antennas and multiple single-antenna users as a mapping function. Generally speaking, the network structure is similar to that of the NOMA-based scenario. In the first stage, an offline learning mechanism is provided for training the DNN. To be specific, the corresponding received signal vectors $\mathbf y$ are obtained according to the values of the fixed channel vector $\mathbf H$ in different directions, and they are used as training samples. Further, we generate the physical DOA $\theta_k$ randomly to constitute a training set with the received signals $\mathbf y$ in the sampling period of $0.01^{\circ}$ within $\{-\frac{\pi}{2}, \frac{\pi}{2}\}$. Concretely, to obtain the training examples, we transmit unit signals to the uplink massive MIMO model in different direction and corresponding received vectors are obtained. It is noted that each group of received vectors are obtained in a specific direction (i.e., each direction is denoted by the corresponding DOA). The training dataset, testing dataset, and validation dataset comprise 150,000, 20000, and 30000 samples, respectively, and the samples in the testing dataset are exclusive of that of the training and validation datasets. After using the SGD with momentum to optimize the network, we apply the offline learning mechanism to deploy the network and perform DOA estimation. To test the performance of the proposed DOA estimation scheme, a fair comparison of the proposed scheme, the spatial basis expansion model (SBEM) method, PA channel tracking mean \cite{pa}, the angle division multiple access (ADMA) user scheduling approach, and the multiple signal classification (MUSIC) method is presented. It can be seen from Fig. \ref{com} that the proposed deep learning-based method outperforms other schemes in terms of DOA estimation. In addition, due to the DNN, not only has a significant enhancement of the mean square error (MSE) performance been achieved in the DOA estimation, but this framework does also not require additional computational complexity.


While the need for a disruptive change in estimation theory appears to be clear, major research efforts have been directed to fulfilling super-resolution channel estimation. Since the history of innovations indicated that dividing the channel estimation into steering vector estimation (i.e., DOA estimation) and complex gain estimation can achieve better performance, as having fewer variables leads to a smaller error, we only perform the complex gain estimation. Similar to the strategy provided for DOA estimation, the complex gain estimation method is proposed, and we can realize channel estimation.



Then, we make a comparison of the performance of the proposed deep learning-based approach with those of the SBEM method and the JSDM scheme. As exhibited in Fig. \ref{channel}, the deep learning-based framework has superior performance in terms of BER compared with the other methods. We conclude that deep learning is an alternative in future communication systems. Inspired by the sparse structure of the massive MIMO, deep learning is capable of fully leveraging its spatial information and improving system performance.

Besides, other state-of-the-art work has provided interesting idea of deep learning-based channel estimation \cite{jin2}. To achieve CSI reduction and reconstruct the CSI for channel estimation, this DNN framework comprises an encoder and an decoder. The encoder transfers the vectors into the codewords with the aid of compressive sensing, and then the CSI can be recovered in the decoder which is developed based on the CNN and the RefineNet. This scheme claims that it can realize low complexity and superior channel estimation performance since it introduces more practical channel distribution and channel feedback.

\subsection*{\centering{Deep Learning-based MmWave Techniques}}

The mmWave technique has been given great importance by the research community since it is an alternative approach for improving spectrum efficiency and facilitating the design of large-scale communication networks. However, there are many challenges in the context of the mmWave technique, such as path loss, penetration loss, and high power consumption. In particular, hybrid precoding has been developed to limit the power consumption and hardware cost of mmWave-based systems, but the existing hybrid precoding methods require bit allocation and have a high computational complexity. We envision that deep learning can optimize the hybrid precoding in an mmWave system, following its remarkable performance when applied to NOMA and massive MIMO. Hence, we integrate deep learning into mmWave massive MIMO and propose a high-efficiency hybrid precoding scheme \cite{mmw}.

\begin{figure*}[!t]
  \centering
  \includegraphics[width=155mm]{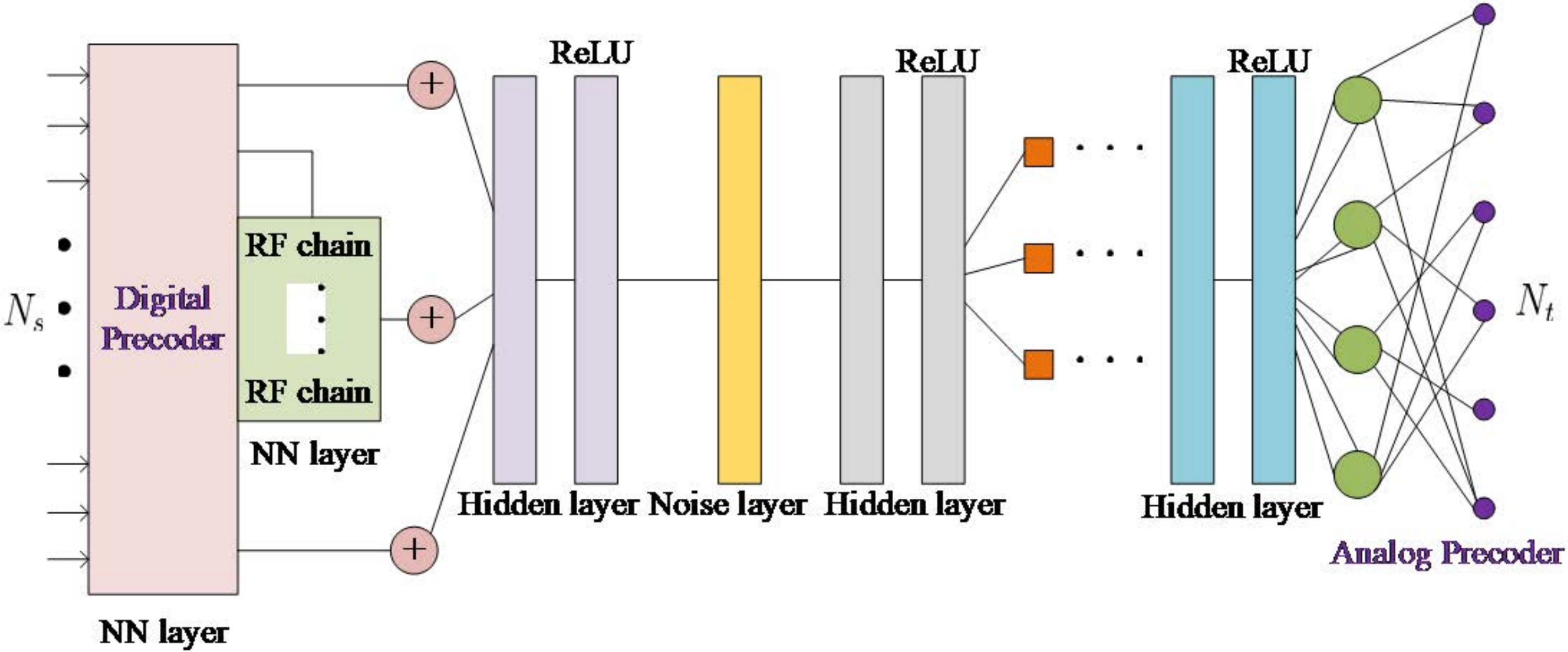}\\
  \caption{Deep learning-based mmWave hybrid precoding framework. Here, $N_s$ represents independent data streams to the user, while $N_t$ denotes the number of the transmitted antennas at the BS.}
  \label{mm}
\end{figure*}

\begin{figure}
  \centering
  \includegraphics[width=87mm]{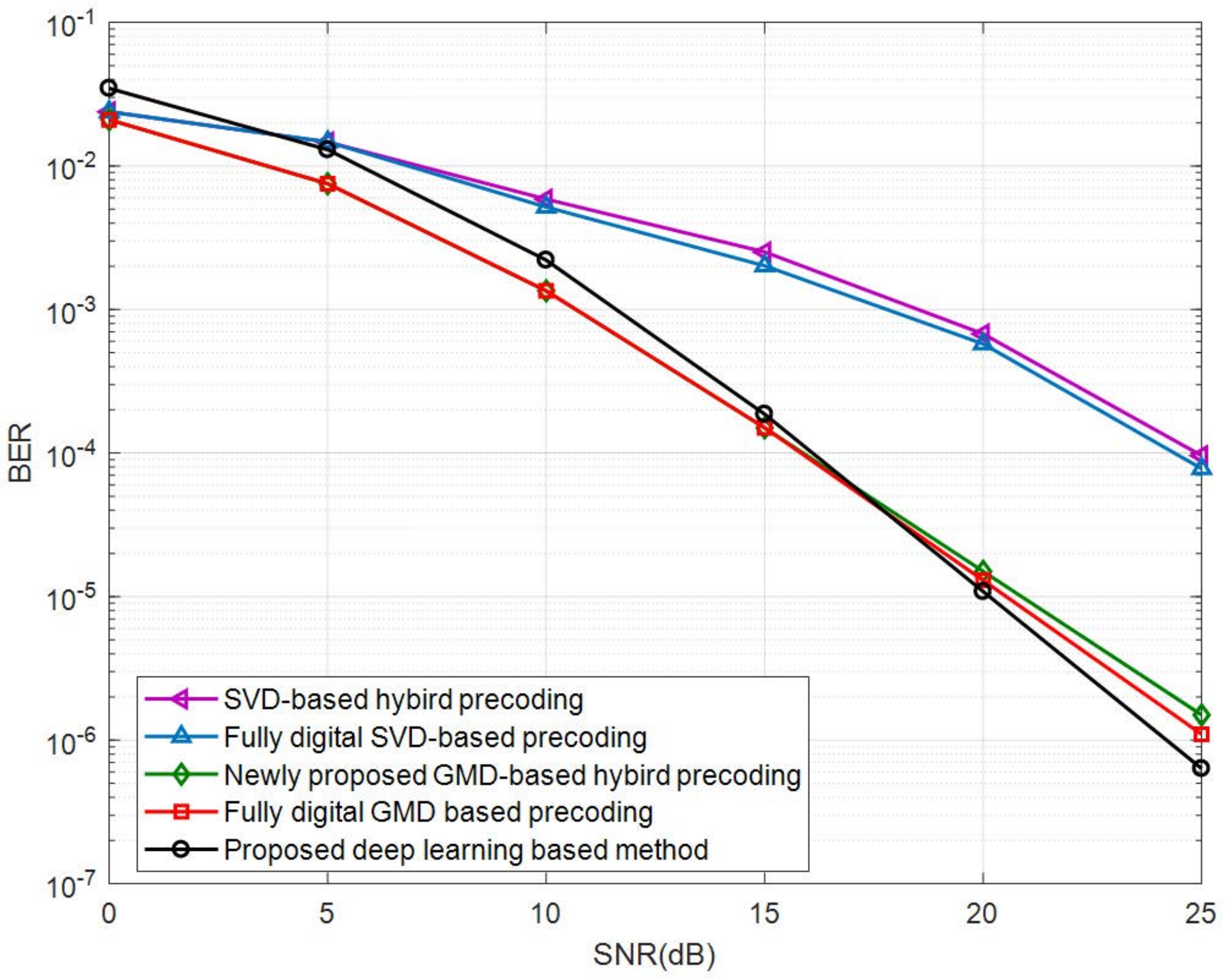}\\
  \caption{Precoding performance of the proposed deep learning-based method, the fully GMD-based method, the newly proposed GMD-based hybrid strategy, the SVD-based hybrid method \cite{3mm}, and the fully digital SVD-based approach.}
  \label{precoding}
\end{figure}

Let us consider a mmWave massive MIMO system, which contains one BS with multi-antennas and many multi-antennas users. To transform the hybrid precoding issue into an optimization problem, we introduce the geometric means decomposition (GMD) to decompose the channel matrix of the mmWave massive MIMO system. Then, in contrast to previous work, we design a DNN to approximate the optimization issue to obtain the best hybrid precoder, as depicted in Fig. \ref{mm}. Specifically, each iteration in the optimization problem is equal to one update process in the training process of the DNN framework. Afterwards, we compare the BER performance of the proposed deep learning-based framework, the fully GMD-based precoding method, the GMD-based hybrid precoding strategy, the SVD-based hybrid precoding method \cite{3mm}, and the fully digital SVD-based precoding approach. Here, we generate the channel statistics based on the Saleh-Valenzuela (SV) channel model at 28 GHz and the angles are generated randomly in the domain of $\{-\frac{\pi}{2}, \frac{\pi}{2}\}$. The ray-tracing simulator \cite{ak} is introduced to generate channel measurements such as the AoAs and the AoDs, and the model is trained for 45k iterations with 0.001 learning rate. Also, momentum of 0.85 and weight decay of 0.0001 are introduced, and there are 500k samples and 20k samples in the training set and the testing set, respectively. We observe from Fig. \ref{precoding} that when the SNR is larger than 16 dB, the deep learning-based framework outperforms other methods. Additionally, it is noted that the computational complexity of the deep learning-based framework is lower than that of other schemes.

Therefore, we conclude that the mmWave massive MIMO system can be optimized using deep learning. The deep learning method can fully extract the sparse features and leverage them to thoroughly learn the whole communication system; this branch of frameworks will be very appealing in next-generation communication networks.

\section*{Future Challenges and Opportunities}

The remarkable capacities of the deep learning technique to deal with enormous data and sharply changing environments indicate that it can improve the performance of the NOMA, the massive MIMO, and the mmWave network. However, many issues have not been resolved, and there are some factors that can overwhelm the system performance. There are numerous challenges to overcome and several new techniques to explore in the future.

\subsection*{\centering{Data Set Acquisition}}

The quality and quantity of the training and testing data set have a great impact on the performance of a deep learning-based framework. In computer science, with the rapid development of natural language processing (NLP), computer vision (CV), and autonomous driving, many well-known and efficient data sets, such as the ImageNet and the MNIST, have been provided. However, in the field of deep learning-based wireless communication, although some data sets such as RML2016 can be employed in some areas, there are few common data set existing in related topics. To facilitate research, it is desirable to create a common and reliable data set for a branch of problems; alternatively, we can develop new software programs that can generate data sets for the corresponding problems in deep learning-based communication networks.

Over the past several decades, various channel models have been exploited, and their performances have been verified. We may test these models in different environments and optimization problems to collect training examples. Additionally, it would be interesting to expand the data set acquisition problem into future research and focus on building data sets similarly to the creation of ImageNet by the Stanford University.

\subsection*{\centering{Models Selection}}

In the context of deep learning-based communication frameworks, the design of neural networks is the core challenge. Many deep learning-based technologies have been developed following general models. For instance, the CNN is always employed for CV, whereas the LSTM is often used in the NLP area. However, we wonder whether there are models for deep learning-based wireless communication; we believe that general models would facilitate the implementation of such frameworks in practice. In engineering projects, not only do the general models improve the convenience of optimizing the communication frameworks, but they can also reduce the cost and time of model selection.
This topic still needs to be extensively explored before useful and highly general models can be obtained.

\subsection*{\centering{Learning Mechanism and Performance Analysis}}

In general, the performance of the deep learning-based communication framework has been demonstrated for channel estimation, encoding and decoding, massive MIMO and other scenarios; however, we have not derived rigorous mathematical proofs and solid theorems to further verify the framework's performance. Furthermore, deriving solid theories would help us understand the communication framework, which is the basis for modifying the network and exploiting more efficient communication frameworks. Yet, since the original input signals are often being transformed into binary signals, one-hot vectors, modulated integers and other styles of data representation for enhancing network performance in the deep learning area, it is not crystal clear that state-of-the-art performance can be obtained in the deep learning-based wireless communication frameworks while changing the styles of data representation. We understand that the rules of learning schemes are still unclear in the field of deep learning-based wireless physical layer, and the optimal results of deep learning-based communication frameworks remain unknown; additionally, we have not yet developed a method for selecting training examples based on such systems. Furthermore, apart from the classical loss function such as the Sigmoid function, the question of whether we can derive a specific function for deep learning-based wireless communication physical layer and enhance its performance remains. Many issues need to be explored further.

\subsection*{\centering{Deep Reinforcement Learning for Wireless Physical Layer}}

Deep reinforcement learning, as an alternative method for solving resource allocation issues, has been proposed in recent years. In 5G, there are numerous resource allocation and energy management problems that need to be addressed, but the current approaches face limitations in such big data processing problems. Hence, the remarkable deep reinforcement learning is capable of optimizing the equipment performance such as CSI, latency, and bandwidth management to tackle the above issues. In particular, this technique is a good candidate for radio resource management by dealing with communication systems that have complex parameters. Thus, to optimize the significant resource management tasks, deep reinforcement learning-based wireless physical layer should be well investigated in the future.

\subsection*{\centering{Models Compression for Deep Learning-based 5G}}

One limitation lies in current deep learning-based frameworks is that some deep learning models for addressing communication issues require high computational complexity, which hinders its implementation on small terminals such as mobile phones. Till now, a large branch of such frameworks are designed based on the LSTM and the CNN. However, these models lead to ultra-high parameters and they cause high complexity both in memory and time. Thus, it indicates that we can design super-performance deep compression schemes to optimize the deep learning-based networks and reduce their complexity, and models compression strategies such as prone, quantization, and Huffman coding can be considered in developing future frameworks.

\balance

\section*{Conclusions}
\label{secconclusion}

In this article, inspired by the state-of-the-art deep learning and 5G communications, we summarize the recent development of deep learning-based wireless physical layers and describe several novel and efficient deep learning-based communication frameworks. In particular, focusing on 5G, three frameworks based on deep learning are presented and their performance are investigated, in which are NOMA, massive MIMO, and mmWave hybrid precoding. However, it must be admitted that many technical implementations are in their infancy with open research questions, and it is a long road ahead to use the deep learning concepts to address wireless physical layer issues thoroughly. Clearly, our investigation of this topic is still rather new, and the complicated DNN hinders current research. For example, the data set acquisition and model selection problems need to be resolved. It is desirable to develop explainable deep learning methods and that we need to build the common data sets which many people support in the field of wireless communication. We hope that our work can provide scholars with a gold mine of entirely new research problems.



\end{document}